\documentclass[10pt,twocolumn,A4]{revtex4-1}

\usepackage{epsfig}
\usepackage{amsfonts}
\usepackage[bbgreekl]{mathbbol}
\usepackage{mathtools}
\usepackage{amsthm}
\usepackage{amsbsy}
\usepackage{multirow}
\usepackage{slashed}
\usepackage{esint}
\usepackage{epsfig}
\usepackage{amsmath,latexsym,amssymb}
\usepackage{graphicx}
\usepackage{psfrag}
\usepackage[latin1]{inputenc}
\usepackage[T1]{fontenc}
\usepackage{subfigure}
\usepackage{nicefrac}
\usepackage{hyperref}

\def\NO{\nonumber}

\newcommand{\be}{\begin{equation}}
\newcommand{\ee}{\end{equation}}

\def\bea{\begin{eqnarray}}
\def\eea{\end{eqnarray}}

\def\beqx{\begin{displaymath}}
\def\eeqx{\end{displaymath}}

\newcommand{\bmat}{\left(\begin{array}}
\newcommand{\emat}{\end{array}\right)}

\newcommand{\ca}{{\cal A}}

\newcommand{\ce}{{\cal E}}
\newcommand{\ck}{{\cal K}}

\newcommand{\cm}{{\cal M}}
\newcommand{\cl}{{\cal L}}
\newcommand{\cf}{{\cal F}}

\newcommand{\co}{{\cal O}}
\newcommand{\cp}{{\cal P}}
\newcommand{\cs}{{\cal S}}
\newcommand{\cq}{{\cal Q}}
\newcommand{\ct}{{\cal T}}

\def\a{\alpha}

\def\d{\delta}
\def\e{\epsilon}
\def\f{\phi}
\def\g{\gamma}

\def\j{\psi}
\def\k{\kappa}
\def\l{\lambda}
\def\m{\mu}
\def\n{\nu}

    \def\om{\omega}
\def\p{\pi}

    \def\th{\theta}
\def\r{\rho}
\def\s{\sigma}

\def\x{\xi}

\def\D{\Delta}

\def\P{\Pi}

\def\bs{\bbalpha}
\def\bs{\bbbeta}
\def\bs{\bbgamma}
\def\bs{\bbdelta}
\def\bs{\bbepsilon}
\def\bs{\bbeta}
\def\bs{\bbsigma}

\def\ca{{\cal A}}

\def\cc{{\cal C}}

\def\ce{{\cal E}}
\def\cf{{\cal F}}

\def\ch{{\cal H}}

\def\ck{{\cal K}}
\def\cl{{\cal L}}
\def\cm{{\cal M}}

\def\co{{\cal O}}
\def\cp{{\cal P}}
\def\cq{{\cal Q}}
\def\car{{\cal R}}
\def\cs{{\cal S}}
\def\ct{{\cal T}}

\def\bb#1{\ensuremath{\mathbb{#1}}} 

\def\bo{{\raise-.3ex\hbox{\large$\Box$}}}               
\def\pa{\partial}                                       
\def\face{{\raise.2ex\hbox{$\displaystyle \bigodot$}\mskip-2.2mu \llap {$\ddot
        \smile$}}}                                   
\def\>{\rangle}                                      
\def\<{\langle}                                      

\newcommand{\sub}[1]{{}_{(#1)}{}}    					 
\def\Hat#1{\widehat{#1}}                             
\def\leftrightarrowfill{$\mathsurround=0pt \mathord\leftarrow \mkern-6mu
        \cleaders\hbox{$\mkern-2mu \mathord- \mkern-2mu$}\hfill
        \mkern-6mu \mathord\rightarrow$}        
\def\dvec#1{\vbox{\ialign{##\crcr
        \leftrightarrowfill\crcr\noalign{\kern-1pt\nointerlineskip}
        $\hfil\displaystyle{#1}\hfil$\crcr}}}           

\def\-{\hphantom{-}}

\begin{document}

\begin{flushright}
\widetext{IFT UAM/CSIC-14-040}
\end{flushright}


\title{Generalized dilatation operator method for non-relativistic holography}

\date{\today}

\author{Wissam Chemissany$^1$}\email{wissam@stanford.edu}
\author{Ioannis Papadimitriou$^2$} \email{ioannis.papadimitriou@csic.es}

\affiliation{$^1$Department of Physics and SITP, Stanford University, Stanford, California 94305 USA\\ $^2$Instituto de F\'isica Te\'orica UAM/CSIC, Universidad Aut\'onoma de Madrid, Madrid 28049, Spain}

\begin{abstract}

We present a general algorithm for constructing the holographic dictionary for Lifshitz and hyperscaling violating Lifshitz backgrounds for any value of the dynamical exponent $z$ and any value of the hyperscaling violation parameter $\theta$ compatible with the null energy condition. The objective of the algorithm is the construction of the general asymptotic solution of the radial Hamilton-Jacobi equation subject to the desired boundary conditions, from which the full dictionary can be subsequently derived. Contrary to the relativistic case, we find that a fully covariant construction of the asymptotic solution for running non-relativistic theories necessitates an expansion in the eigenfunctions of two commuting operators instead of one. This provides a covariant but non-relativistic grading of the expansion, according to the number of time derivatives.

\end{abstract}
 
\maketitle

{\bf Introduction} In recent years, great effort has been devoted to the use of holographic models in order to gain a deeper understanding of the strong coupling physics in condensed matter systems. The gauge/gravity duality has proven an instrumental tool in studying the strongly coupled dynamics near quantum critical points
exhibiting Lifshitz \cite{Kachru:2008yh,Taylor:2008tg}  or Schr\"odinger \cite{Balasubramanian:2008dm,Son:2008ye} symmetry. 
More recently, gravity duals to non-relativistic systems that  transform non-trivially under scale transformations have been put forward \cite{Charmousis:2010zz,Huijse:2011ef,Dong:2012se,Shaghoulian:2011aa}. The geometries dual to such hyperscaling violating Lifshitz (hvLf) quantum systems are of the form
\be\label{thetaz}
ds_{d+2}^2 =\frac{  du^2 - u^{-2(z-1)} dt^2 + d\vec x^2 }{\ell^{-2} u^{2(d-\theta)/d} },
\ee
where $d$ is the spatial dimension, $z$ and $\theta $ are respectively the Lifshitz and hyperscaling violation exponents, and $\ell$ is the Lifshitz radius. This metric transforms non-trivially under scale transformations as 
\be
\label{scalecovariance}
\vec x \to \lambda \vec x
,\;
t \to \lambda^{z} t
,\;
u \to \lambda u
,\;
ds_{d+2}^{2} \to \l^{\frac{2\th}{d}} ds_{d+2}^{2}.
\ee
By computing the energy of supergravity fluctuations around the background (\ref{thetaz}) one can unambiguously determine the location of the ultraviolet (UV) of the dual quantum field theory, corresponding to the conformal boundary of the geometry (\ref{thetaz}), to be at $u\to 0$, independently of the value of the exponents $z$ and $\th$ \cite{Peet:1998wn,Dong:2012se}. The only restriction we shall impose on the exponents $z$ and $\th$ is the null energy condition, which leads to seven distinct cases for the values of $z$ and $\th$ \cite{hvLf}. However, the only two solutions that allow for $z<1$ require $\th > d+z$, in which case the on-shell action is UV finite and as a result there are no well defined Fefferman-Graham asymptotic expansions \cite{hvLf}. The marginal case $\th=d+z$ requires separate analysis. Our discussion here and in \cite{hvLf} therefore focuses on the case $z>1$.

For earlier work on asymptotically Lifshitz backgrounds, their hyperscaling violating versions and various string theory embeddings we refer the reader to the following recent papers and references therein \cite{Donos:2010tu,Gath:2012pg,Gouteraux:2012yr,Bueno:2012vx,Narayan:2012wn}. The literature primarily relevant to us here though concerns earlier work on holographic renormalization and the holographic dictionary for asymptotically Lifshitz backgrounds. In particular, holography for the Einstein-Proca theory with Lifshitz boundary conditions has been discussed from a bottom up perspective in  \cite{Ross:2009ar,Horava:2009vy,Ross:2011gu,Mann:2011hg,Baggio:2011cp,Baggio:2011ha,Griffin:2011xs,Griffin:2012qx}, while in \cite{Cassani:2011sv,Chemissany:2011mb,Chemissany:2012du,Christensen:2013lma,Christensen:2013rfa,Korovin:2013bua,Korovin:2013nha} AdS embeddings (or limits) of Lifshitz backgrounds were utilized in order to deduce the non-relativistic dictionary from the relativistic one in special cases. 

Our aim in this letter and the accompanying main paper \cite{hvLf} is to present a general algorithm for the construction of the holographic dictionary of non-relativistic theories, that can be applied to theories with or without a UV fixed point and for any value of the dynamical exponents that is consistent with the null energy condition. The main tool for deriving the holographic dictionary, which includes the the Fefferman-Graham expansions, the identification of the sources and the dual operators, as well as the boundary counterterms required to render the variational problem well posed, is a general covariant asymptotic solution of the radial Hamilton-Jacobi (HJ) equation. Our main result is a general and efficient algorithm for the recursive solution of the HJ equation, based on the covariant expansion in eigenfunctions of two commuting operators, which are related to 
the dilatation operator \cite{Papadimitriou:2004ap} and its generalization for running theories \cite{Papadimitriou:2011qb}.

{\bf The Model} We consider the class of theories defined by the action
\begin{widetext}
\be\label{action}
S_\x=\frac{1}{2\k^2}\int_\cm d^{d+2}x\sqrt{-g}e^{d\x \f}\left(R
-\a_\x (\pa\f)^2-Z_\x F^2-W_\x B^{2}-V_\x\right)
+\frac{1}{2\k^2}\int_{\pa\cm} d^{d+1}x\sqrt{-\g}2e^{d\x\f}K,
\ee
\end{widetext}
where $\a_\x$ and $\x$ are arbitrary parameters and $Z_\x(\f)$, $W_\x(\f)$ and $V_\x(\f)$ are unspecified functions of the real scalar field $\f$. In order to maintain the $U(1)$ gauge symmetry in the presence of a mass for the vector field we have introduced a St\"uckelberg field $\om$ so that $B_\m=A_\m-\pa_\m\om$ is gauge invariant. The parameter $\x$ has been introduced to allow us to interpret our results in any desired Weyl frame. In particular, the $\x$ dependence of (\ref{action}) follows from the Weyl transformation $g\to e^{2\x\f}g$ of the $\x=0$ (Einstein frame) action. Under such a transformation the various parameters and functions of the action transform as $\a_\x=\a-d(d+1)\x^2$, $Z_\x(\f)=e^{-2\x\f}Z(\f)$,
$W_\x(\f)=W(\f)$, and $V_\x(\f)=e^{2\x\f}V(\f)$,
where quantities without the subscript $\x$ refer to the Einstein frame. The advantage of keeping $\x$ arbitrary in our analysis is that we can impose Lifshitz boundary conditions in a generic $\x$ frame and cover both Lifshitz and hvLf boundary conditions in the Einstein frame \cite{hvLf}. In the relativistic case, $z=1$, this trick has been employed in the study of holography for non-conformal branes \cite{Kanitscheider:2008kd}.

{\bf Lifshitz \& Hyperscaling Violating Lifshitz} The action (\ref{action}) admits asymptotically locally Lifshitz solutions of the form 
\be\label{Lifshitz-solution} 
ds^{2}=d r^2-e^{2 z r} dt^{2}+e^{2r} d\vec{x}^2,\; B=\frac{\cq  e^{\e r}}{\e Z_o} dt,\; \f=\m r,
\ee
if $ 
V_{\xi}\sim V_o e^{2  (\r +\xi ) \f}$,  $Z_{\xi}\sim Z_o e^{-2  (\xi +\n )\f}$, and $W_{\xi}\sim W_o e^{2 \s \f}$ asymptotically,
where the parameters of the theory, $V_o$, $Z_o$, $W_o$, $\n$, $\r$, $\s$, $\a_\x$, $\x$, are related to the Lifshitz boundary condition parameter $z$ and the integration constants $\m$, $\e$ and $\cq$ as 
\bea
   && \rho=-\xi,\; \n=-\x+\frac{\e-z}{\m},\;\s=\frac{z-\e}{\m},\;\cq^2=\frac12Z_o(z-1)\e,\NO\\
   &&\e=\frac{(\a_\x+d^2\x^2)\m^2-d\m\x+z(z-1)}{z-1},\NO\\
   && W_o= 2 Z_o \e (d + z + d \m\x  - \e),\NO\\
   && V_o=-d (1 + \m \x) (d + z + d \m \x) - (z - 1) \e.
\eea
In practice, the parameters we choose to specify at will are $z>1$, $\a>0$, $Z_o>0$,  $\x$ and $\m$. In the Einstein frame (\ref{Lifshitz-solution}) are hvLf solutions with $\th=-d\x\m$ and are equivalent to the solutions presented in \cite{Gouteraux:2012yr}.
 
{\bf Radial Hamiltonian Formalism} Our starting point for the derivation of the full holographic dictionary is the radial Hamiltonian formalism for the action (\ref{action}), where the radial coordinate plays the role of Hamiltonian `time'. Decomposing the bulk fields as 
\bea
&&ds^2=(N^2+N_iN^i)dr^2+2N_idr dx^i+\g_{ij}dx^idx^j,\NO\\
&&A=A_r dr+A_idx^i,
\eea
the Hamiltonian takes the form 
\be
H=\int d^{d+1}x\left(N\ch+N_i\ch^i+A_r\cf\right),
\ee 
where $N$, $N_i$ and $A_r$ are Lagrange multipliers imposing the Hamiltonian, momentum and $U(1)$ gauge constraints, respectively
\begin{widetext}
\begin{align}
\label{constraints}
\begin{aligned}
&0=\ch=-\frac{\k^2}{\sqrt{-\g}}e^{-d\x\f}\left\{2\left(\p^{ij}\p_{ij}-\frac1d\p^2\right)+\frac{1}{2\a}\left(\p_\f-2\x\p\right)^2
+\frac14Z^{-1}_\x(\f)\p^i\p_i+\frac12W^{-1}_\x(\f)\p_\om^2\right\}\\
&\rule{0.8cm}{0cm}+\frac{\sqrt{-\g}}{2\k^2}e^{d\x\f}\left(-R[\g]+\a_\x\pa^i\f\pa_i\f+Z_\x(\f)F^{ij}F_{ij}+W_\x(\f)B^iB_i+V_\x(\f)\right),\\
&0=\ch^i=-2D_j\p^{ji}+F^i{}_j\p^j+\p_\f\pa^i\f-B^i\p_\om,\quad 0=\cf=-D_i\p^i+\p_\om.
\end{aligned}
\end{align}
\end{widetext} 
These constraints provide a full description of the dynamics in the Hamilton-Jacobi (HJ) formalism -- there is no need to use the second order equations of motion. This is achieved by expressing the canonical momenta in two different ways. Firstly, they are written as gradients of the Hamilton's principal function $\cs[\g,A,\f,\om]$ as
\be\label{HJ-momenta}
\p^{ij}=\frac{\d \cs}{\d\g_{ij}},\quad \p^i=\frac{\d\cs}{\d A_i},\quad \p_\f=\frac{\d\cs}{\d\f},\quad \p_\om=\frac{\d\cs}{\d\om},
\ee
and the constraints (\ref{constraints}) are interpreted as functional partial differential equations (PDEs) for $\cs[\g,A,\f,\om]$. However, only the Hamiltonian constraint provides a non-trivial dynamical equation. The momentum constraint simply requires that $\cs$ be invariant with respect to diffeomorphisms on the radial slice, while the $U(1)$ constraint implies that $\cs$ depends on $A_i$ and $\om$ only through the gauge-invariant field $B_i$. Once a complete integral of these PDEs is known, equating the gradients (\ref{HJ-momenta}) with the standard expressions for the momenta in terms of the velocities leads to first order flow equations that can be integrated to obtain the full radial dependence of the fields. We refer to \cite{hvLf} for the full set of flow equations for our model. 

The radial Hamiltonian formulation of the dynamics is particularly suited for developing the holographic dictionary, both for asymptotically AdS and non AdS backgrounds. The radial coordinate plays the role of energy scale in the dual field theory, and so it is natural that it is singled out. From the point of view of the bulk theory, the presence of the boundary naturally gives rise to a Gaussian normal radial coordinate. 
Moreover, the functional $\cs$ is precisely the on-shell action, which is interpreted holographically as the generating function of connected correlation functions. The long distance divergences of the on-shell action correspond to a certain asymptotic solution of the HJ equation. The boundary term required to remove these divergences can be defined in terms of such an asymptotic solution of the HJ equation \cite{Boer:1999xf,Papadimitriou:2010as}. The same boundary term, both for asymptotically AdS and non AdS backgrounds, ensures that the variational problem is well posed \cite{Papadimitriou:2005ii,Papadimitriou:2010as}. Moreover, the integration functions parameterizing the symplectic space of asymptotic solutions are 
automatically organized into sources and 1-point functions in the Hamiltonian formalism, the latter appearing as integration functions parameterizing a complete integral of the HJ equation, while the former emerging as integration functions of the first order flow equations. More generally, given the asymptotic solution of the HJ equation, the flow equations can be integrated  to obtain the Fefferman-Graham expansions. Multi-scale expansions,  where e.g. every power is dressed with an infinite series of logs as for Improved Holographic QCD \cite{Papadimitriou:2011qb}, can be handled  much more efficiently than directly solving asymptotically the second order equations of motion.   

Our main objective, therefore, is to develop a general algorithm for solving the HJ equations asymptotically for the class of theories (\ref{action}) and for Lifshitz asymptotics with any $z>1$. The full holographic dictionary can then be constructed using this asymptotic solution \cite{hvLf}.

{\bf Superpotential vs Boundary Conditions } The first order flow equations relate the leading term of the asymptotic solution of the HJ equation to the leading asymptotic form of the fields. In order for unconstrained sources to appear there must be no conditions involving transverse derivatives on the leading terms of the asymptotic expansions. This implies that the leading asymptotic form of the fields must follow from a zero derivative solution of the HJ equation, i.e. from a `superpotential'. In the presence of a massive vector field, diffeomorphism and gauge invariance dictate that the most general zero derivative solution of the HJ equation takes the form  
\be\label{HJ-zero-order-solution}
\cs\sub{0}=\frac{1}{\k^2}\int d^{d+1}x\sqrt{-\g}U(\f,B^2),
\ee
for some superpotential $U(\f,B^2)$. Using this ansatz for Hamilton's principal function in the flow equations and demanding that the resulting asymptotic form of the metric be asymptotically locally Lifshitz not only determines the asymptotic form of $U(\f,B^2)$ and its first derivatives, but also imposes a (second class) constraint on the asymptotic form of the vector field $B_i$, namely  
\be\label{Lifshitz-constraint} 
B_i\sim B_{oi}=\sqrt{-Y_o(\f)} \;\bb n_i, 
\ee
where $\bb n_i$ is the unit normal to the constant time surfaces and $Y_o(\f):=-(z-1)/2\e Z_\x(\f)$, so that $B^2_o=Y_o(\f)$. Moreover, denoting $\f=: X$ and $B^2=:Y$, the superpotential and its derivatives must take the asymptotic form
\begin{align}
\label{U-asymptotics}
\begin{aligned}
& U(X,Y_o(X))\sim e^{d\x X}\left(d(1+\m\x)+z-1\right),\\
& U_Y(X,Y_o(X))\sim -\e e^{d\x X}Z_\x(X),\\
& U_X(X,Y_o(X))\sim e^{d\x X}\left(-\m\a_\x+d\x(d+z)\right).
\end{aligned} 
\end{align}
To determine the full superpotential we need to solve the PDE resulting from inserting (\ref{HJ-zero-order-solution}) in the HJ equation, subject to these asymptotic conditions. 
The resulting superpotential generically is a Taylor expansion in $Y-Y_o$, or equivalently $B_i-B_{oi}$. The field $Y-Y_o$ sources a scalar operator which can be relevant, marginally relevant, or irrelevant depending on the values of the various parameters defining the model. When the dual operator is relevant then only a finite number of terms in the Taylor expansion are required to obtain an asymptotic complete integral, while in the marginally relevant case the full Taylor expansion is required. If the dual operator is irrelevant then the source of $Y-Y_o$ must be set to zero to preserve the Lifshitz boundary conditions.

{\bf Recursive Solution of the HJ Equation} Given the leading solution (\ref{HJ-zero-order-solution}) of the HJ equation, we seek to determine the subleading terms in the form of a covariant expansion in the eigenfunctions of a suitable operator. There are two essential requirements for this operator. Firstly, the covariant expansion in eigenfunctions of this operator must be compatible with the radial asymptotic expansion in the sense that terms in the covariant expansion are also radially subleading relative to the preceding terms. Secondly, (\ref{HJ-zero-order-solution}) must be an eigenfunction for any superpotential $U$. This requirement is necessary in order to apply the algorithm to models dual to theories with running couplings in the UV, such as non-conformal branes, holographic QCD \cite{Papadimitriou:2011qb}, or hvLf asymptotics in the present case. For boundary conditions corresponding to a UV fixed point, relativistic or not, all fields have definite scaling dimensions and one can use the dilatation operator \cite{Papadimitriou:2004ap}. In the presence of running couplings though we need an operator that is blind to these couplings, such as the generalized dilatation operator introduced in \cite{Papadimitriou:2011qb}. Of course, the generalized dilatation operator can be applied to theories with UV fixed points as well -- the resulting expansion will simply be a particular resummation of the one obtained with the usual dilatation operator. 

There are in fact two operators that satisfy these conditions for (\ref{HJ-zero-order-solution}), namely     
\bea
&&\Hat\d:= \int d^{d+1}x\left(2\g_{ij}\frac{\d}{\d\g_{ij}}+B_i\frac{\d}{\d B_i}\right),\NO\\
&&\d_B:=\int d^{d+1}x\left(2Y^{-1}B_iB_j\frac{\d}{\d\g_{ij}}+B_i\frac{\d}{\d B_i}\right).
\eea
Indeed, for any $U$, $\Hat\d \cs\sub{0}= (d+1) \cs\sub{0}$,  $\d_B \cs\sub{0}= \cs\sub{0}$.
Moreover, these operators commute with each other, which means that we can construct simultaneous eigenfunctions of both. The eigenfunctions of $\Hat\d$ are easy to understand. Any local covariant quantity that contains a fixed power of the induced metric $\g_{ij}$ and the vector $B_i$, as well as their covariant derivatives, is an eigenfunction of $\Hat\d$. Since any function of the quantity $B^2$ -- not just powers -- is an eigenfunction, however, covariance implies that only the factors of $\g_{ij}$ and $B_i$ that are contracted with transverse derivatives count in determining the eigenvalue of $\Hat\d$. In fact, it can be shown that $\Hat\d$ counts the number of transverse derivatives. Namely, a covariant local functional $\cs\sub{2k}$ containing $2k$ derivatives is an eigenfunction of $\Hat\d$ with eigenvalue $d+1-2k$, where $d+1$ is the contribution of the volume element.  

The structure of the eigenfunctions of $\d_B$ can be understood using the fact that $\d_B$ annihilates the projection operator $\s^i_j:=\d^i_j-Y^{-1}B^iB_j$, i.e. $\d_B\s^{ij}=0$. An eigenfunction $\cs\sub{2k}$ of $\Hat\d$ with $2k$ derivatives can be split into a sum of up to $k+1$ terms containing respectively $0, 1, \ldots, k$ factors of $\s^{ij}$. This is achieved systematically as follows. Terms in which all $2k$ derivatives are contracted with $B^i$ are eigenfunctions of $\d_B$ with eigenvalue $1-2k$, since every factor of $B^i$ contributes $-1$ to the eigenvalue and the $1$ comes from the volume element. Terms where $2k-2$ derivatives are contracted with $B^i$ and $2$ derivatives are contracted with $\g^{ij}$ are not eigenfunctions of $\d_B$ but they can be written as a sum of two eigenfunctions of $\d_B$ with eigenvalues respectively $1-2(k-1)$ and $1-2k$ by writing $\g^{ij}=\s^{ij} + Y^{-1}B^iB^j$. This process can be repeated for all terms with $2k$ derivatives in order to split $\cs\sub{2k}$ into a sum of eigenfunctions of $\d_B$ with eigenvalues $1-2\ell$, $\ell=0,1,\ldots,k$. 

This procedure provides a {\em grading} of any eigenfunction of $\Hat\d$ and allows us to look for a solution of the HJ equation in the form of a double covariant expansion in simultaneous eigenfunctions of $\Hat\d$ and $\d_B$, namely
\be\label{covariant-expansion-s}
\cs=\sum_{k=0}^\infty\cs\sub{2k}=\sum_{k=0}^\infty\sum_{\ell=0}^k\cs\sub{2k,2\ell},
\ee
where $\Hat\d\cs\sub{2k,2\ell}=(d+1-2k)\cs\sub{2k,2\ell}$ and $\d_B\cs\sub{2k,2\ell}=(1-2\ell)\cs\sub{2k,2\ell}$. This grading becomes especially meaningful in the context of Lifshitz boundary conditions. The constraint (\ref{Lifshitz-constraint}) imposed by such boundary conditions forces $B_i$ to be asymptotically aligned with the unit normal, $\bb n_i$, to the constant time slices. This in turn implies that the projection operator $\s_{ij}$, asymptotes to the spatial metric $\bs_{ij}:=\g_{ij}+\bb n_i\bb n_j$, and the operator $\d_B$ counts time derivatives. 

Our algorithm for solving the HJ equation beyond the leading order solution (\ref{HJ-zero-order-solution}) therefore involves two expansions. Firstly, a graded covariant expansion of the form (\ref{covariant-expansion-s}) in simultaneous eigenfunctions of $\Hat\d$ and $\d_B$, and secondly a functional Taylor expansion in $B_i-B_{oi}$ around the second class constraint (\ref{Lifshitz-constraint}) imposed by the Lifshitz boundary conditions. In particular, every term in (\ref{covariant-expansion-s}) admits an expansion in $B_i-B_{oi}$. Inserting this multiple expansion of $\cs$ in the HJ equation leads to a tower of recursion relations in $k,\ell$ for every order in the Taylor expansion in $B-B_o$. Writing   
\be
\cs\sub{2k,2\ell}=\int d^{d+1}x\cl\sub{2k,2\ell},
\ee  
the recursion relations for the zeroth order in the $B-B_{oi}$ expansion take the form \cite{hvLf}
\be\label{kth-order-0}
\ck^{-1}(\f)\left(\frac{\d}{\d\f}\int d^{d+1}{x'}
-\cc_{k,\ell}\ca'(\f)\right)\cl^0_{(2k,2\ell)}=\car^0_{(2k,2\ell)},
\ee
where $\cc_{k,\ell}=(d+1-2k)+(z-1)(1-2\ell)$ and the inhomogeneous term $\car^0_{(2k,2\ell)}$ is determined by the 2-derivative terms of the HJ equation at order $k=1$, and by the lower order momenta for order $k>1$. Moreover, 
\be
e^{\ca(\f)}=Z_\x^{-\frac{1}{2(\e-z)}}\sim e^{\f/\m},
\ee
and the kernel $\ck(\f)\sim 1/\m$ can also be expressed in terms of the potentials of the model. Similar recursion relations can be derived for the higher orders in the $B-B_{oi}$ expansion \cite{hvLf}. These recursion relations are the key element in our iterative algorithm. By expanding the solution of the HJ equation in a graded covariant expansion in eigenfunctions of the operators $\Hat\d$ and $\d_B$ and in a functional Taylor expansion around the second class constraint imposed by Lifshitz asymptotics we have reduced the problem of solving the HJ equation at each order  in the $B_i-B_{oi}$ expansion to the recursion problem of the relativistic case studied in \cite{Papadimitriou:2011qb}. 
  
The solution of the recursion relations (\ref{kth-order-0}) depends crucially on the parameter $\m$, which determines whether the scalar is asymptotically running ($\m\neq 0$) or asymptotically constant ($\m=0$). The latter case corresponds to a UV fixed point and the recursion relations become {\em algebraic}, providing an extremely efficient algorithm for constructing the holographic dictionary for the Einstein-Proca theory \cite{Ross:2011gu,Mann:2011hg,Griffin:2011xs,Baggio:2011cp,Baggio:2011ha}, or the Einstein-Proca-Scalar model studied in \cite{Chemissany:2012du,Christensen:2013lma,Christensen:2013rfa}, including the Lifshitz conformal anomalies.  When $\m\neq 0$ the functional integration over the scalar can be carried out using the method developed in \cite{Papadimitriou:2011qb}. This is necessary to study holography for hvLf backgrounds since the hyperscaling violating parameter $\th$ in the Einstein frame is proportional to $\m$. However, when the various potentials defining the model are exactly exponentials the recursion relations become effectively algebraic even for $\m\neq 0$ \cite{hvLf}.  

The iterative procedure using the recursion relation (\ref{kth-order-0}) for the $\co(B-B_o)^0$ solution of the HJ equation need only be carried out as long as $\cc_{k,\ell}-d\m\x\leq 0$, or equivalently $(d+1-2k)+(z-1)(1-2\ell)-\th\leq 0$. At order $\co(B-B_o)^m$ in the Taylor expansion the corresponding inequality takes the form $(d+1-2k)+(z-1)(1-2\ell)-\th-m\D_-\leq 0$, where $\D_+:=d+z+\th-\D_-$ is the dimension of the scalar operator dual to the mode $Y-Y_o$ \cite{hvLf}. The sum of all terms $\cs_{(2k,2\ell)}^m$ for which this inequality holds are UV divergent and can be identified with (minus) the local covariant boundary counterterms required to regularize the variational problem and the on-shell action \cite{hvLf}.
If the values of $z$, $\th$ and $\D_-$ are such that there exist non-negative integers $k$ and $0\leq \ell\leq k$ saturating this inequality, then the corresponding term $\cs_{(2k,2\ell)}^0$ will have a pole which needs to be regularized by introducing explicit cut-off dependence, as is well known from the relativistic case. When $\m=0$ such terms give rise to non-relativistic conformal anomalies, but in the presence of a running dilaton ($\m\neq 0$) the corresponding logarithmic divergences can absorbed into the dilaton, thus avoiding any explicit dependence on the cut-off \cite{hvLf} and hence the appearance of a conformal anomaly. It is important to note that, in contrast to the relativistic case, terms with different number of spatial and time derivatives can contribute to the non-relativistic conformal anomaly, as has been observed before for the $d=z=2$ Einstein-Proca theory \cite{Baggio:2011ha}. 

Moreover, there is always an independent solution of the HJ equation starting with a term $\Hat\cs_{reg}$ that has dilatation weight zero, and hence is UV finite \cite{hvLf}. This term can be parameterized as 
\be\label{Sreg}
\Hat\cs_{reg}=\int d^{d+1}x\left(\g_{ij}\Hat\p^{ij}+B_i\Hat\p^i+\f\Hat\p_\f\right),
\ee 
where $\Hat\p^{ij}$, $\Hat\p^i$ and $\Hat\p_\f$ are undetermined integration functions, only subject to the momentum constraint in (\ref{constraints}). The term $\Hat\cs_{reg}$ containing the integration functions is required in order to have an asymptotic complete integral of the Hamilton-Jacobi equation. The holographic dictionary identifies $\Hat\cs_{reg}$ with the regularized generating function of correlation functions in the dual quantum field theory, and the integration functions $\Hat\p^{ij}$, $\Hat\p^i$ and $\Hat\p_\f$ are related to the regularized one-point functions of local operators \cite{hvLf}.  

{\bf Asymptotic Expansions \& Ward Identities} Having obtained the asymptotic solution of the HJ equation in the form of a covariant graded expansion in eigenfunctions of the operators $\Hat\d$ and $\d_B$ Taylor expanded in $B_i-B_{oi}$, the Fefferman-Graham expansions are determined by integrating the first order flow equations. To account for the anisotropic scaling of the metric and the gauge field due to the Lifshitz boundary conditions we parameterize the induced metric and the gauge field as
\bea\label{decomposition}
&&\g_{ij}dx^idx^j=-(n^2-n_a n^a)dt^2+2n_adtdx^a+\s_{ab}dx^a dx^b,\NO\\
&&A_idx^i=a dt+ A_a dx^a,
\eea
where $a,b$ run over the spatial directions only. The flow equations determine the leading asymptotic behavior of the anisotropic fields to be \cite{hvLf}
\bea\label{sources} 
&& n\sim e^{zr}n\sub{0}(x),\;
n_a\sim e^{2r}n\sub{0}_a(x),\;
\s_{ab}\sim e^{2r}\s\sub{0}_{ab}(x),\NO\\
&&\f\sim \m r+\f\sub{0}(x), \; \j\sim e^{-\D_-r}\j_-(x),\;\om\sim\om\sub{0}(x),
\eea
where $\j:=Y_o^{-1}B_o^i(B_i-B_{oi})$ and $n\sub{0}(x)$, $n\sub{0}_a(x)$, $\s\sub{0}_{ab}(x)$, $\om\sub{0}(x)$, $\f\sub{0}(x)$ and $\j_-(x)$ are arbitrary sources. Note that there is no independent source for the gauge field allowed by the Lifshitz boundary conditions since its asymptotic form is completely determined by the rest of the fields according to 
\be
A_i\sim \sqrt{\frac{z-1}{2\e Z_o}}\;n\sub{0}e^{\frac{(\e-z)\f\sub{0}}{\m}}e^{\e r}\d_{it}\left(1+e^{-\D_-r}\j_-\right)+\pa_i\om\sub{0}. 
\ee
This is a direct consequence of the second class constraint (\ref{Lifshitz-constraint}) and is in agreement with what has been found in the vielbein formalism \cite{Ross:2011gu}. Note also that the source $\om\sub{0}(x)$ corresponds to a pure gauge transformation and so it does not source an independent operator. 

The modes conjugate to these sources, i.e. the renormalized 1-point functions, correspond to certain combinations of the integration functions $\Hat\p^{ij}$, $\Hat\p^i$ and $\Hat\p_\f$ in (\ref{Sreg}). Defining the linear combinations  
\bea\label{vevs-1}
\Hat\ct^{ij}&:=&-\frac{e^{-d\x\f}}{\sqrt{-\g}}\left(2\Hat\p^{ij}+Y_o^{-1}B_o^iB_o^jB_{ok}\Hat\p^k\right),\NO\\
\Hat\co_\f &:=&\frac{e^{-d\x\f}}{\sqrt{-\g}}\left(\Hat\p_\f+(\n+\x)B_{oi}\Hat\p^i\right),\NO\\
\Hat\co_\j&:=&\frac{e^{-d\x\f}}{\sqrt{-\g}}B_{oi}\Hat\p^i, \quad \Hat\ce^i:=  \frac{e^{-d\x\f}}{\sqrt{-\g}}\sqrt{-Y_o}\bs^i_j\Hat\p^j,
\eea
(\ref{Sreg}) leads to the following source-1-point function pairs \cite{hvLf}: 
\[\begin{array}{|l|l|}
	\hline \hline 
	\hspace{0.7in}\mbox{1-point function}& \hspace{0.18in}\mbox{source}\quad\\
	\hline
	 \quad\Hat\P^i_j:=\bs^i_k\bs_{jl}\Hat\ct^{kl}\sim e^{-(d+z-\th)r}\P^i_j(x)\quad & \quad \s_{(0)ab}\\
	  \quad\Hat\cp^i:=-\bs^i_k\bb n_l\Hat\ct^{kl}\sim e^{-(d+2-\th)r}\cp^i(x)\quad& \quad n_{(0)a}\\
	 \quad\Hat\ce:=-\bb n_k\bb n_l\Hat\ct^{kl}\sim e^{-(d+z-\th)r}\ce(x)\quad& \quad n_{(0)}\\
	  \quad\Hat\ce^i\sim e^{-(d+2z-\th)r}\ce^i(x)\quad  &\quad 0\\
	\quad\Hat\co_\f\sim  e^{-(d+z-\th)r}\co_\f(x)\quad & \quad\f_{(0)}\\
	\quad\Hat\co_\j\sim  e^{-\D_+r}\co_\j(x)\quad & \quad \j_-\\
	\hline\hline
\end{array}\]
The modes $\P^i_j(x)$, $\cp^i(x)$, $\ce(x)$, $\ce^i(x)$ are respectively the spatial stress tensor, momentum density, energy density and energy flux, which comprise the energy-momentum complex of a non-relativistic field theory \cite{Ross:2011gu}. In agreement with \cite{Ross:2011gu}, the energy flux is an irrelevant operator and its source is set to zero by the Lifshitz boundary conditions. Inserting these 1-point functions in the momentum constraint for $\Hat\p^{ij}$, $\Hat\p^i$ and $\Hat\p_\f$ leads to the non-relativistic diffeomorphism Ward identities \cite{hvLf}, which for flat boundary take the form
\bea
 	&&\bb D_j\Hat\P^i_i+\bb n^jD_j\Hat\cp_i+\Hat\co_\f\bb D_i\f+\Hat\co_\j\bb D_i\j=0,\NO\\
 	&&\bb n^iD_i\Hat\ce+\bb D_i\Hat\ce^i+\Hat\co_\f\bb n^iD_i\f=0,\quad \bb D_i\Hat\cp^i=0,
\eea
where $\bb D_i$ denotes the covariant derivative with respect to the boundary metric $\bs_{ij}$. Moreover, the transformation of (\ref{Sreg}) under infinitesimal anisotropic Weyl transformations gives rise to the trace Ward identity  
\be
	z\Hat\ce+\Hat\P^i_i+\D_-\j\Hat\co_\j=\left\{\begin{matrix}\m\Hat\co_\f,& \m\neq 0,\\
	\ca,& \m=0,\end{matrix}\right.		
\ee
where $\ca$ is the conformal anomaly \cite{hvLf}.

{\bf Summary \& Conclusions}
 We present a general recursive algorithm for solving asymptotically the radial Hamilton-Jacobi equation for an Einstein-Proca-scalar theory with arbitrary scalar couplings, from which the full holographic dictionary is obtained. Lifshitz and hyperscaling violating Lifshitz  asymptotics are imposed covariantly as constraints  corresponding to turning off the source for the energy flux, which is an irrelevant operator. The asymptotic solution of the Hamilton-Jacobi equation takes the form of a covariant expansion in eigenfunctions of two commuting operators,  which provide a generalization of the dilatation operator \cite{Papadimitriou:2004ap} to anisotropic and scale covariant boundary conditions. The full details of the algorithm, together with the explicit form of the asymptotic expansions and the holographic dictionary for a number of concrete examples are presented in the accompanying paper \cite{hvLf}.

\begin{acknowledgements}
We thank Blaise Gout\'eraux, Jelle Hartong, Cynthia Keeler and Simon Ross for useful discussions. IP would also like to thank the Asian Pacific Center for Theoretical Physics (APCTP) as well as the Centro de Ciencias de Benasque Pedro Pascual for the hospitality during the completion of this work. WC is supported by the SITP, Stanford and the Arab Fund for Economic and Social Development. The work of IP is funded by the Consejo Superior de Investigaciones Cient\'ificas and the European Social Fund under the contract JAEDOC068. This work has also been supported by the ESF Holograv Programme, the Spanish Ministry of Economy and Competitiveness under grant FPA2012-32828, Consolider-CPAN (CSD2007-00042), the Spanish MINECO's ``Centro de Excelencia Severo Ochoa'' Programme under grant SEV-2012-0249, as well as by the grant HEPHACOS-S2009/ESP1473 from the C.A. de Madrid.
\end{acknowledgements}


\bibliography{hvLf}
\bibliographystyle{h-physrev}
\end{document}